\documentclass[12pt]{iopart}
\usepackage{graphicx,amssymb,subfigure}

\usepackage{iopams}  
\begin{document}

\title[$q$-deformed optical tomograms and estimation of quadrature moments]{Janus-faced tomograms and retrieval of quadrature moments for $q$-deformed states}

\author[SK, CS]{S. Kannan$^{1}$ and C. Sudheesh$ ^2 $}
\address{$ ^1 $ Department of Quantum Science and Technology, Research School of Physics, The Australian National University,
Canberra, ACT 2601, Australia.\\$ ^2 $ Department of Physics, Indian Institute of Space Science and Technology, Thiruvananthapuram 695 547, India.}
\ead{$ ^1 $Kannan.S@anu.edu.au, $ ^2 $sudheesh@iist.ac.in}
\vspace{10pt}
\date{\today}
\begin{abstract}
In this work, we derive the optical tomograms of various $q$-deformed quantum states. 
We found that the optical tomograms of the states under consideration exhibit a fascinating `Janus faced' nature, irrespective of the deformation parameter $q$. We also derived a general method to extract the quadrature moments from the optical tomograms of any $q$-deformed states. We also note that this technique can be used in high-precision experiments to observe deviations from the standard quantum mechanical behavior.   
\end{abstract}
%
%
%
%
%

\section{Introduction}\label{sec1}
The quantum state provides all the information of a given system. Usually, this is as a vector ($\vert{\psi}\rangle$) for pure states and a density matrix ($\rho$) for mixed states. In an experiment, if one is unaware of whether the system is pure or mixed, many sets of measurements are required to reconstruct the density matrix. This is briefly the method of quantum state tomography\,\cite{lvovsky2009continuous}. Optical tomography was introduced in quantum optics by reconstructing  Wigner distribution and the density matrix of squeezed states of light\,\cite{smithey1993measurement}. Optical tomography is based on the finding that the rotated quadrature phase can be expressed in  quasiprobability distributions and vice versa\,\cite{vogel1989determination}. Thus an optical tomogram of a quantum system contains all the information encoded in the system's density matrix. The study of the physical properties of quantum states and the estimation of errors in the experimentally observed values can be evaluated using an optical tomogram\,\cite{bellini2012towards}. In recent literature, optical tomograms were used to study various nonclassical properties\,\cite{rohith2016signatures,sharmila2017signatures},   quantum mechanical evolution of states\,\cite{rohith2015visualizing}, and also as an indicator of the nonclassicality in a quantum system\,\cite{rohith2023homodyne}. Optical tomograms were also used to study $q$-deformed coherent states\,\cite{jayakrishnan2017q} and string coherent states\,\cite{wani2022construction}. 

Deformed quantum mechanics can play a vital role in the search for possible new physics at a high energy scale. The general feature of such spaces is that they are non-commutative and have a well-defined mathematical structure from quantum group symmetries. $q$-deformation in quantum mechanics is widely studied, ranging from the nonclassical features of $q$-deformed states\,\cite{dey2015q,dey2016noncommutative,berrada2019noncommutative,kannan2022squeezing}, a study of cosmic microwave background radiation\,\cite{zeng2017thermal}, in loop quantum gravity to construct observables with cosmological constants\,\cite{dupuis2014observables}, the realization of quasibosons\,\cite{gavrilik2011quasibosons}, etc. In this work, we apply the concept of optical tomograms to $q$-deformed quantum states. We derive the $q$-deformed versions of well-known nonclassical states and an expression to calculate quadrature moments from the optical tomograms of any $q$-deformed states. We note that this can be used to derive various nonclassical features of deformed quantum states and can be used in the search for new physics.

The paper is organized as follows: section 2  compares the optical tomogram of $q$-deformed `Janus faced' partner states. We introduce $q$-deformed squeezed vacuum and excited states as the eigenstates of $q$-deformed two-photon annihilation operators. In section 3, the optical tomogram is expressed as an expectation value of normal ordered power series of $q$-deformed annihilation and creation operators.   We also derive  an equation to obtain moments from the optical tomogram of any $q$-deformed state. Section 4 is devoted to the conclusion.

\section{Janus-faced tomograms}\label{sec2}
It is known that $q$-deformed states, such as the $q$-deformed cat and squeezed states, are nonclassical. While analyzing some of these $q$-deformed states, we noticed that
some of their optical tomograms reveal a ‘Janus-faced’ nature. We found that  this property of quantum states was earlier reported in non-deformed scenarios\,\cite{laha2018estimation}. This section analyzes two  ‘Janus-faced’ optical tomogram pairs: (i). $q$-deformed even cat state and $q$-deformed squeezed vacuum state, and (ii). $q$-deformed odd cat state and $q$-deformed squeezed excited state.\\
Let us commence our study by considering a one-dimensional $ q $-deformed oscillator satisfying the algebra 
\begin{equation}\label{eq1}
	AA^{\dagger}-q^{2}A^{\dagger}A=1, \textnormal{ }\vert{q}\vert<1,
\end{equation} where $ A $ and $ A^{\dagger} $ are the $ q $-deformed annihilation and creation operators$ \, $\cite{arik1976hilbert}. In literature, this kind of deformation is called the math-type $ q $-deformation.  Using $ q $-deformed integers $ [n]_{q} $, the algebra in Eq.$ \, $\ref{eq1} can be defined on the $ q $-deformed Fock space such that the following relations are satisfied:
\begin{equation}\label{eq2}
	A\vert{n}\rangle_{q}=\sqrt{[n]_{q}}\vert{n-1}\rangle_{q},\textnormal{ }A^{\dagger}\vert{n}\rangle_{q}=\sqrt{[n+1]_{q}}\vert{n+1}\rangle_{q},
\end{equation}
where the $ q $-deformed integer $ [n]_{q} $ and $ \vert{n}\rangle_{q} $ is defined as
\begin{equation}\label{eq3}
	[n]_{q}=\frac{1-q^{2n}}{1-q^{2}}\textnormal{, } \vert{n}\rangle_{q}=\frac{A^{\dagger n}}{\sqrt{[n]_{q}!}}\vert{0}_{q}\textnormal{ and }[n]_{q}!=\prod_{k=1}^{n}[k]_{q}.
\end{equation}
The algebra described above, which deviates from the non-deformed version, is an example of non-standard quantum mechanical behavior. In the limit $q\rightarrow1$, everything converges to the non-deformed scenario satisfying the algebra $ [a,a^{\dagger}]=1 $.

The nonclassical properties of $ q $-deformed cat states are well documented in the literature\,\cite{dey2015q}. In terms of $ q $-deformed coherent state $ \vert{\alpha}\rangle_{q}\,$\cite{dey2013time}, these are given by
\begin{equation}\label{eq20}
	\vert{\alpha,\pm}\rangle_{q}=\frac{1}{N_{q}(\alpha,\pm)}\left(\vert{\alpha}\rangle_{q}\pm\vert{-\alpha}\rangle_{q}\right),
\end{equation}
where $ \vert{\alpha,+}\rangle_{q} $ and $ \vert{\alpha,-}\rangle_{q} $ represents the even and odd $ q $-deformed cat states with the normalization constant $ N_{q}(\alpha,\pm) $. The `Janus-faced' $ q $-deformed partner states of the even and odd cat states are the eigenstates of the $ q $-deformed version of two-photon annihilation operators$ \, $\cite{mehta1992eigenstates}, $ A^{\dagger-1}A $ and $ AA^{\dagger-1} $.  For the non-deformed scenario, the eigenstates of the two-photon annihilation operators$ \, $(TAO's) are analogous to the squeezed vacuum state and the squeezed first excited state$ \, $\cite{mehta1992eigenstates}.
From Eq.\,\ref{eq2}, we obtain the action of TAO's on the $ q $-deformed number state for $ n\geq2 $
\begin{equation}\label{eq21}
	A^{\dagger-1}A\vert{n}\rangle_{q}=\sqrt{\frac{[n]_{q}}{[n-1]_{q}}}\vert{n-2}\rangle_{q},\textnormal{ }AA^{\dagger-1}\vert{n}\rangle_{q}=\sqrt{\frac{[n-1]_{q}}{[n]_{q}}}\vert{n-2}\rangle_{q}.
\end{equation}
Now let us calculate the eigenstates for these operators. Let $ \vert\xi\rangle_{q} $ be the right eigenstate of $ A^{\dagger-1}A $. In the deformed Fock basis, $ \vert\xi\rangle_{q} $ can be expanded as 
\begin{equation}\label{eq22}
	\vert\xi\rangle_{q}=\sum_{n=0}^{\infty}C_{n}\vert n\rangle_{q}.	
\end{equation}
The action of the first TAO on $ \vert\xi\rangle_{q} $ gives
\begin{equation}\label{eq23}
	A^{\dagger-1}A\vert\xi\rangle_{q}=\sum_{n=0}^{\infty}C_{n+2}\sqrt{\frac{[n+2]_{q}}{[n+1]_{q}}}\vert n\rangle_{q}=\xi\vert\xi\rangle_{q}.
\end{equation}
We can see that the coefficient $ C_{n} $ satisfies the recurrence relation 
\begin{equation}\label{eq}
	\xi C_{n}=\sqrt{\frac{[n+2]_{q}}{[n+1]_{q}}}C_{n+2}.
\end{equation}
For even $ n $, we can see that the state
\begin{equation}\label{eq25}
	\vert\xi\rangle_{q}=N\sum_{n=0}^{\infty}\xi^{n}\sqrt{\frac{[2n-1]_{q}!!}{[2n]_{q}!!}}\vert 2n\rangle_{q},
\end{equation}
where $ N $ is the normalization factor and for $ \xi=-e^{i\theta}\tanh(r) $, $ \vert\xi\rangle_{q} $ becomes the $ q $-deformed squeezed vacuum state$ \, $\cite{noormandipour2014f}. Similarly, we can compute the right eigenstate of the second TAO, and for odd $ n $, the eigenstate $ \vert\xi_{1}\rangle_{q} $ is
\begin{equation}\label{eq26}
	\vert\xi_{1}\rangle_{q}=N_{1}\sum_{n=0}^{\infty}\xi_{1}^{n}\sqrt{\frac{[2n+1]_{q}!!}{[2n]_{q}!!}}\vert 2n+1\rangle_{q}.
\end{equation}
Here $ N_{1} $ is the normalization constant and for $ \xi_{1}=-e^{i\theta}\tanh(r) $, we obtain the $ q $-deformed squeezed first excited state.

\begin{figure}[t]
	\centering
	\subfigure[]{\includegraphics[width=0.455\linewidth]{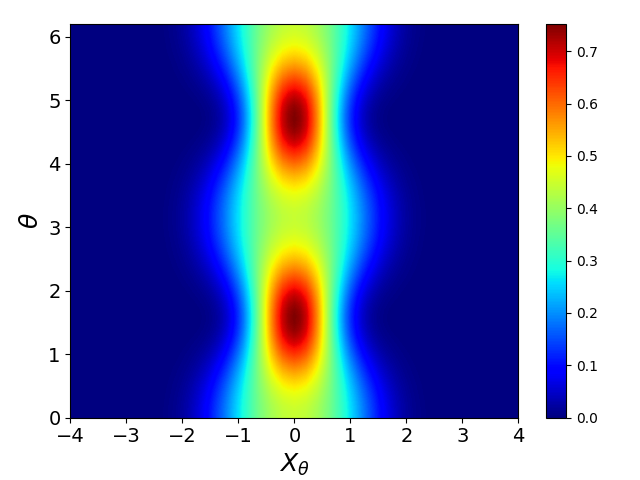}}
	\subfigure[]{\includegraphics[width=0.45\linewidth]{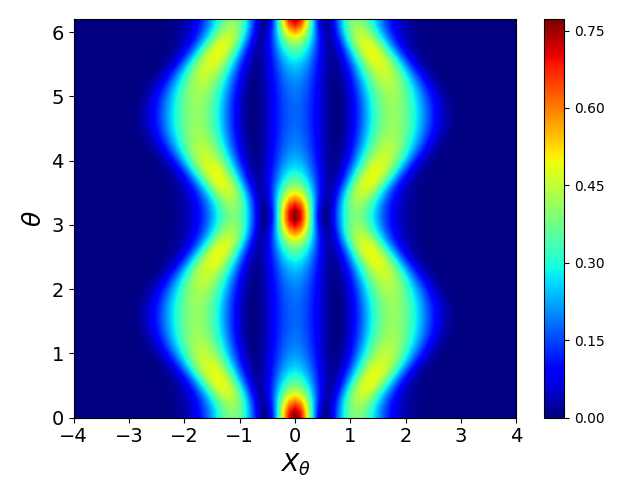}}
	\subfigure[]{\includegraphics[width=0.455\linewidth]{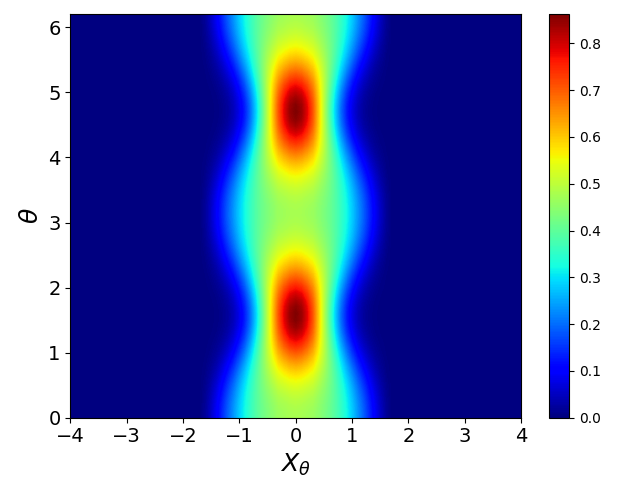}}
	\subfigure[]{\includegraphics[width=0.455\linewidth]{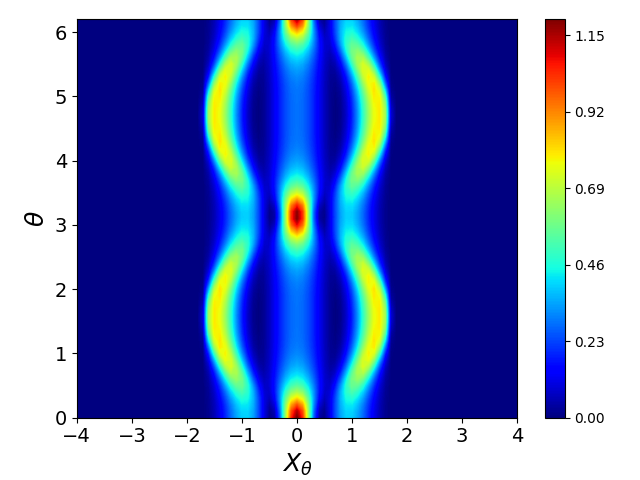}}
	\caption{$ q$-deformed optical tomogram for $ q $-even cat state ((a) and (c)) and $ q $-squeezed vacuum state ((b) and (d)) with $ \vert\alpha\vert^2=r=0.5 $ for $ q=0.9 $\,((a) and (b)) and $ q=0.7 $\,((c) and (d)).}\label{fig1}
	\centering
\end{figure}

\begin{figure}[ht]
	\centering
	\subfigure[]{\includegraphics[width=0.45\linewidth]{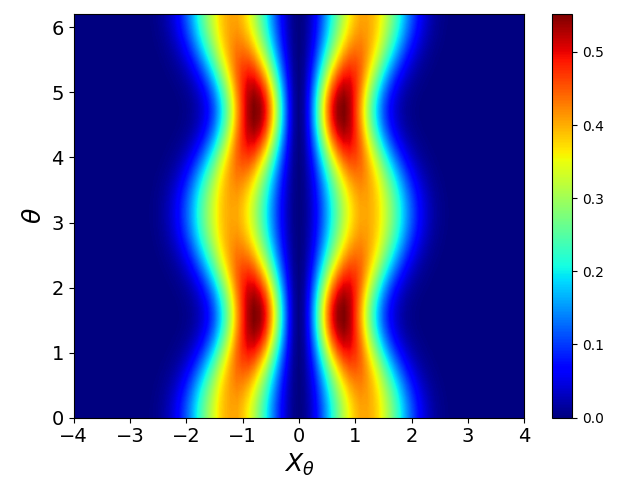}}
	\subfigure[]{\includegraphics[width=0.45\linewidth]{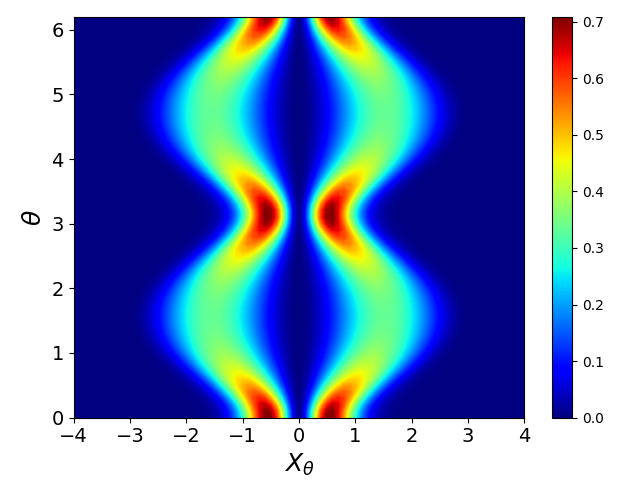}}
	\subfigure[]{\includegraphics[width=0.45\linewidth]{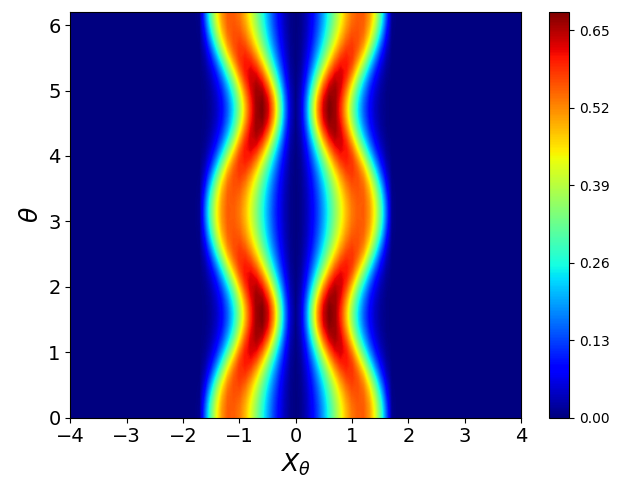}}
	\subfigure[]{\includegraphics[width=0.45\linewidth]{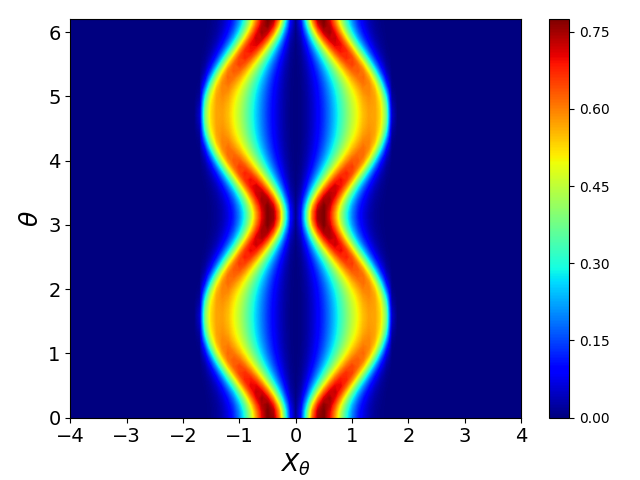}}
	\caption{$ q$-deformed optical tomogram for $ q $-odd cat state ((a) and (c)) and $ q $-squeezed excited state ((b) and (d)) with $ \vert\alpha\vert^2=r=0.5 $ for $ q=0.9 $\,((a) and (b)) and $ q=0.7 $\,((c) and (d)).}\label{fig2}
	\centering
\end{figure}

Figs.\,\ref{fig1} and \ref{fig2} are the $ q$-deformed tomograms (explained in the next section) for the states defined above. We have taken the parameters $ \alpha $, $ \xi $, and $ \xi_{1} $ to be real. These tomograms reveal their `Janus-faced' nature. Interestingly, the `Janus-faced' tomogram pairs appear the same except for a phase difference of $ \pi/2 $. These properties exist for all $ q $ values. Also, with an increase in deformation (small $q$ value), the strands in the tomographic plane become narrower and denser (large value of $\omega(X_{\theta},\theta)$). 

\section{Estimation of moments from optical tomogram}\label{sec3}

Expressing density operator $ \rho $ using normal ordered moments $( \langle a^{\dagger\alpha}a^{\beta}\rangle) $ for non-deformed algebra was derived in$ \, $\cite{wunsche1996tomographic,wunsche1990reconstruction}. Here we will derive the $ q $-deformed version of the same.

Let us consider the expansion of an operator $ F $ as a normal ordered power series in $ A $ and $ A^{\dagger} $
\begin{equation}\label{eq4}
	F=\sum_{\alpha,\beta=0}^{\infty}F_{\alpha,\beta}A^{\dagger \alpha}A^{\beta}.
\end{equation}
Using Eq.$ \, $\ref{eq2}, it can be easily shown that \begin{equation}\label{eq5}
	_{q}\langle{m}|F|{n}\rangle_{q}=\sum_{r=0}^{min(\alpha,\beta)}F_{\alpha-r,\beta-r}\left(\frac{[m]_{q}![n]_{q}!}{[r]_{q}!^{2}}\right)^{1/2}.
\end{equation}
Now, we can guess the expression for $ F_{\alpha,\beta} $:
\begin{equation}\label{eq6}
	F_{\alpha,\beta}=\sum_{k=0}^{min(\alpha,\beta)}\frac{(-1)^{k}q^{k(k-1)}}{[k]_{q}!\sqrt{[\alpha-k]_{q}![\beta-k]_{q}!}}  {_{q}\langle}{\alpha-k}|F|{\beta-k}\rangle_{q}.
\end{equation}
This can be proved as follows:
	\begin{eqnarray}\label{eq7}
	&{ }{ }	\sum_{k=0}^{min(\alpha,\beta)}\frac{(-1)^{k}q^{k(k-1)}{_{q}\langle}{\alpha-k}|F|{\beta-k}\rangle_{q}}{[k]_{q}!\sqrt{[\alpha-k]_{q}![\beta-k]_{q}!}}\nonumber  \\&\hspace*{-20mm}=\sum_{k=0}^{min(\alpha,\beta)}\sum_{l=0}^{min(\alpha-k,\beta-k)}\hspace*{-5mm}\frac{(-1)^{k}q^{k(k-1)}}{[k]_{q}!\sqrt{[\alpha-k]_{q}![\beta-k]_{q}!}}\left(\frac{[\alpha-k]_{q}![\beta-k]_{q}!}{[l]_{q}!^{2}}\right)^{1/2}F_{\alpha-k-l,\beta-k-l}.
	\end{eqnarray}
	Using the $ q  $-binomial formula$ \, $\cite{si1994q}, we can express
	\begin{equation}\label{eq8}
		\delta_{p,0}=(1-1)^{p}_{q}=\sum_{k=0}^{p}\frac{[p]_{q}!}{[k]_{q}![p-k]_{q}!}(-1)^{k}q^{k(k-1)}.
	\end{equation}
	Inserting Eq.$ \, $\ref{eq8} into RHS of Eq.$ \, $\ref{eq7} along with the change of variable $ l=p-k $, we obtain
	\begin{equation}\label{eq9}
		\sum_{p=0}^{min(i,j)}\sum_{k=0}^{p}\frac{(-1)^{k}q^{k(k-1)}}{[k]_{q}![p-k]_{q}!}F_{\alpha-p,\beta-p}=F_{\alpha,\beta},
	\end{equation}
	which concludes our proof.
$ F $ can also be expressed as
\begin{equation}\label{eq10}
	F=\sum_{m,n=0}^{\infty}|{n}\rangle_{q}{_{q}\langle{m}}|Tr\{|{m}\rangle_{q}{_{q}\langle{n}|}F\}.
\end{equation}
Now by applying Eq.$ \, $\ref{eq6} for $ |{m}\rangle_{q}{_{q}\langle{n}|} $, we have
	\begin{eqnarray}\label{eq11}
		F&{ }{ }=\sum_{m,n=0}^{\infty}\frac{|{n}\rangle_{q}{_{q}\langle{m}}|}{\sqrt{[m]_{q}![n]_{q}!}}\sum_{k=0}^{\infty}\frac{(-1)^{k}q^{k(k-1)}}{[k]_{q}!}Tr\{A^{\dagger m+k}A^{n+k}F\}\nonumber\\&=\sum_{\alpha,\beta=0}^{\infty}\sum_{k=0}^{min(\alpha,\beta)}\frac{(-1)^{k}q^{k(k-1)}|{\beta-k}\rangle_{q}{_{q}\langle{\alpha-k}}|}{[k]_{q}!\sqrt{[\alpha-k]_{q}![\beta-k]_{q}!}}Tr\{A^{\dagger \alpha}A^{\beta}F\}.
	\end{eqnarray}
Thus using Eq.$ \, $\ref{eq11}, we can express the density operator $ \rho $ in terms of moments of $ q $-deformed annihilation and creation operator as
\begin{eqnarray}\label{eq12}
		&\rho=\sum_{\alpha,\beta=0}^{\infty}\rho(\alpha,\beta){_{q}\langle A^{\dagger \alpha}A^{\beta}\rangle_{q}},\textnormal{ where}\\&\rho(\alpha,\beta)=\sum_{k=0}^{min(\alpha,\beta)}\frac{(-1)^{k}q^{k(k-1)}|{\beta-k}\rangle_{q}{_{q}\langle{\alpha-k}}|}{[k]_{q}!\sqrt{[\alpha-k]_{q}![\beta-k]_{q}!}}.
\end{eqnarray}
Using the homodyne $ q $-deformed quadrature $ X_{\theta}\,$\cite{jayakrishnan2017q}, we can compute the optical tomogram $ \omega(X_{\theta},\theta) $ for the density matrix $ \rho $ as\,\cite{vogel1989determination}
\begin{equation}\label{eq13}
	\omega(X_{\theta},\theta)={_{q}\langle X_{\theta}}|\rho|X_{\theta}\rangle_{q}.
\end{equation}
After inserting Eq.$ \, $\ref{eq12} into the above equation, we have
\begin{equation}\label{eq14}
	\omega(X_{\theta},\theta)=\sum_{\alpha,\beta=0}^{\infty}{_{q}\langle X_{\theta}}|\rho(\alpha,\beta)|X_{\theta}\rangle_{q}{_{q}\langle}A^{\dagger \alpha}A^{\beta}\rangle_{q}.
\end{equation}
With the quadrature representation of $ q $-deformed Fock state, we compute
\begin{eqnarray}\label{eq15}
		{_{q}\langle X_{\theta}}|\rho(\alpha,\beta)|X_{\theta}\rangle_{q}=&\sum_{k=0}^{min(\alpha,\beta)}\frac{(-1)^{k}q^{k(k-1)}}{[k]_{q}!\sqrt{[\alpha-k]_{q}![\beta-k]_{q}!}}\nonumber\\&\times J_{{\alpha-k}_{q}}(X_{\theta})J_{{\beta-k}_{q}}(X_{\theta})e^{i(\alpha-\beta)\theta}\vert{\Psi_{0_{q}}(X_{\theta})}\vert^{2},
\end{eqnarray}  
where $ J_{n_{q}}(X_{\theta}) $ is an $ n^{th} $ degree polynomial that satisfies the three-term recurrence relation$ \, $\cite{jayakrishnan2017q}
\begin{equation}\label{eq16}
	J_{n+1_{q}}(X_{\theta})=\frac{1}{\sqrt{[n+1]_{q}}}\left[\frac{2X_{\theta}}{\sqrt{1+q^{2}}}J_{n_{q}}(X_{\theta})-\sqrt{[n]_{q}}J_{n-1_{q}}(X_{\theta})\right].
\end{equation}
Here, $ \Psi_{0_{q}}(X_{\theta}) $ is the wavefunction for the $ q $-deformed vacuum state. We can show that for finite sums over the polynomial $ J_{n_{q}} $, the following identity is satisfied:
\begin{equation}\label{eq17}
	\sum_{k=0}^{min(\alpha,\beta)}\frac{(-1)^{k}q^{k(k-1)}}{[k]_{q}!\sqrt{[\alpha-k]_{q}![\beta-k]_{q}!}}J_{{\alpha-k}_{q}}J_{{\beta-k}_{q}}=\frac{\sqrt{[\alpha+\beta]_{q}!}}{[\alpha]_{q}![\beta]_{q}!}J_{{\alpha+\beta}_{q}}.
\end{equation}
Now by inserting Eq.$ \, $\ref{eq17} into Eq.$ \, $\ref{eq15}, Eq.$ \, $\ref{eq14} can be expressed as
\begin{equation}\label{eq18}
	\omega(X_{\theta},\theta)=\sum_{\alpha,\beta=0}^{\infty}\vert{\Psi_{0_{q}}(X_{\theta})}\vert^{2}e^{i(\alpha-\beta)\theta}\frac{\sqrt{[\alpha+\beta]_{q}!}}{[\alpha]_{q}![\beta]_{q}!}J_{{\alpha+\beta}_{q}}(X_{\theta}){_{q}\langle}A^{\dagger \alpha}A^{\beta}\rangle_{q}.
\end{equation}
As the polynomial $ J_{n_{q}} $ satisfies the three-term recurrence relation$ \, $(Eq.$ \, $\ref{eq16}), Favard's theorem guarantees its orthogonality.  Now, by multiplying Eq.$ \, $\ref{eq18} with $ J_{{\gamma}_{q}}(X_{\theta})$$ \,(\gamma=\alpha+\beta) $, integration over the variable $ X_{\theta} $, one obtains a linear combination of normally ordered moments ${_{q}\langle}A^{\dagger \alpha}A^{\gamma-\alpha}\rangle_{q} $. Using a little algebra, one can find all the normally ordered moments. For example, let us elaborate on the calculation for $ {_{q}\langle}A\rangle_{q}$. Consider two arbitrary  different angles $ \theta_{1} $, $ \theta_{2} $, $\theta_{1}-\theta_{2}\neq\pi $. Now, one can obtain ${_{q}\langle}A\rangle_{q}$ by performing the following operation on the optical tomogram $ \omega(X_{\theta},\theta) $.
\begin{equation}\label{eq19}
	\hspace*{-10mm}{_{q}\langle}A\rangle_{q}=\frac{1}{2i\sin(\theta_{2}-\theta_{1})}\int_{-\infty}^{\infty}dX_{\theta}\big(e^{i\theta_{2}}\omega(X_{\theta},\theta_{1})-e^{i\theta_{1}}\omega(X_{\theta},\theta_{2})\big)J_{1_{q}}(X_{\theta}).	
\end{equation} 
The optical tomogram can be used to calculate the squeezing and higher-order squeezing for $ q $-deformed states.  Nonclassical properties like these can be obtained from instantaneous tomograms for a system evolving in time. 
\section{Conclusion}
In this paper, we analyzed the optical tomograms of $q$-deformed versions of nonclassical states. The optical tomograms of these states are computed for different deformation parameters. We arranged the states into `Janus faced' partners, and this holds for all deformation values. We also found that with a decrease in the value of $q$ (increase in deformation), the strands in the tomographic plane became thin; also, the maximum value of $\omega(X_{\theta},\theta)$ increased.

We note that the optical tomograms are well suited to study various nonclassical properties of quantum states. We derived a method to extract the quadrature moments from the optical tomogram of deformed states, and this is useful in calculating nonclassical properties such as quadrature squeezing. Also,  this technique can be applied to tabletop experiments where the deviation from standard quantum mechanical behavior is to be tested.
\section*{Data availability statement}
All data supporting this study's findings are included in the article. 
\section*{References}
\bibliographystyle{iopart-num}
\bibliography{reference1}
\end{document}